\documentclass
  [prc,twocolumn,twoside,showpacs,amsmath,amsfonts,amssymb]{revtex4}

\newcommand\etal{\textit{~et~al.}}
\newcommand\Je{J_\text{e}}
\newcommand\Jd{J_\text{d}}
\newcommand\Sp{\text{\itshape Sp}}
\newcommand\bra\langle
\newcommand\ket\rangle
\newcommand\vc\mathbf

\usepackage{graphicx}

\begin{document}

\title{Cooperation of anti-aligning and aligning shell-model forces
  for $N=Z$}

\author{K. Neerg\aa rd}

\affiliation{Fjordtoften 17, 4700 N\ae stved, Denmark}

\email{kai@kaineergard.dk}

\begin{abstract}
  For two neutrons and two protons or two neutron holes and two proton
  holes in a single $j$-shell, the state $|\phi\ket$ with isospin and
  seniority zero and the lowest angular momentum zero state
  $|\chi\ket$ produced by an attractive interaction of quasinucleon
  pairs with angular momentum $2j$ have a large overlap for all
  relevant $j$ and large contents of quasinucleon pairs with angular
  momenta $2j$ and~0, respectively. In the $1f_{7/2}$ and $1g_{9/2}$
  shells, the large negative matrix elements of the effective
  interaction in these two channels relative to most of the rest
  therefore \emph{cooperate} to produce a ground state which is
  essentially a linear combination of $|\phi\ket$ and $|\chi\ket$ with
  comparable coefficients. Interaction matrix elements in other
  channels influence significantly the \emph{ratio} of these
  coefficients. The state $|\phi\ket$ makes up about 80\% of the
  calculated ground states. The overlaps of the latter with
  $|\chi\ket$ are \emph{less} in the $1g_{9/2}$ shell than in the
  $1f_{7/2}$ shell.
\end{abstract}


\maketitle

\section{Background}

In a recent article~\cite{ref:Qi}, Qi\etal\ discussed the ground-state
structure of nuclei with two neutrons and two protons or two neutron
holes and two proton holes in spherical shells with high
single-nucleon angular momentum $j$ such as the shells $1f_{7/2}$,
$1g_{9/2}$, and $1h_{11/2}$. The authors compare contributions of the
configurations
\begin{equation}
  ((j_1j_2)\Je(j_3j_4)\Je)0
\end{equation}
with $\Je=0$ (where `e' stands for `equal') and
\begin{equation}
  ((j_1j_3)\Jd(j_2j_4)\Jd)0
\end{equation}
with $\Jd=2j$ (where `d' stands for `different') in a vector-coupling
notation with the total magnetic quantum number suppressed. The first
and second quasinucleons are the neutrons or neutron holes and the
third and fourth quasinucleons the protons or proton holes, and all
$j_i$ are equal to $j$. In single-$j$-shell calculations with
interactions taken from experimental data and the classic analysis by
Schiffer and True~\cite{ref:SchTr}, they find that $\Je=0$ contributes
51--62~\% whereas $\Jd=2j$ contributes 92--95~\%. From these and other
results of calculations they infer that in $N=Z$ nuclei approaching
$^{100}$Sn a spin-aligned isoscalar pair mode replaces as the dominant
coupling scheme the isovector pairing mode prevalent in the bulk of
the chart of nuclides.

\section{State with isospin and seniority zero}

\begin{figure}
{\center\includegraphics[width=\columnwidth]{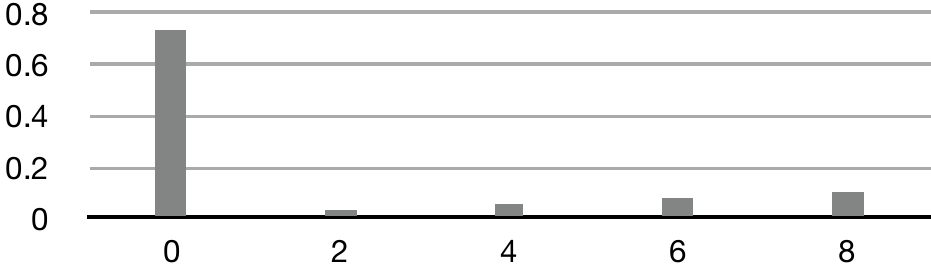}\par}
  \caption{\label{fig:0} $\Je$ distribution in $|\phi\ket$ for
    $j=9/2$.}
\end{figure}

For a deeper analysis of these observations the Pauli principle and
conservation of isospin must be taken into account. Such an analysis
is presented in the following. Throughout, $J_e$ is assumed to be
even, as required by the Pauli principle. The states $|\Je\ket$ thus
span the angular momentum $I=0$ space. While isospin is conserved by
the interactions considered by Qi\etal\ and the calculated ground
states have isospin $T=0$, the individual $|\Je\ket$ are not
eigenstates of $T$ for $j\ge3/2$. The $I=0$ space has a
$\lceil2j/3\rceil$-dimensional $T=0$ subspace. If finite-dimensional,
that is, for $j\ge3/2$, the orthogonal subspace has
$T=2$~\cite{ref:Qi2}. It can be shown by Flowers's
method~\cite{ref:Fl} that in each such finite-dimensional maximal
eigenspace of $T$ within the $I=0$ space, a one-dimensional subspace
carries the seniority $s=0$ representation of the symplectic group
$\Sp(2j+1)$, while the orthogonal subspace, if finite-dimensional,
belongs to the representation $s=4$, $t=T$, where $t$ is the reduced
isospin. Since $|\Je=0\ket$ is symplectically invariant, it belongs to
the $s=0$ subspace (two-dimensional for $j\ge3/2$) of the $I=0$ space.
Therefore, in each finite-dimensional maximal eigenspace of $T$ within
the $I=0$ space the state $|s=0\ket$ is obtained by projecting
$|\Je=0\ket$ onto that space and thus has the maximal content of
$\Je=0$. Figure~\ref{fig:0}
shows for $j=9/2$ the distribution of $\Je$ in the $s=T=0$ state
$|\phi\ket$. It is seen that $\Je=0$ makes up only 73\% of it.
Therefore 51--62~\% of $\Je=0$ in the calculated ground states is
equivalent to $|\phi\ket$ contributing 70--85~\%. A very similar
picture emerges for other $j$.

It is well known that a general isobarically invariant
two-quasinucleon interaction can be written as $V=\sum_J V_J$ with
\begin{equation}
  V_J=c_J\sum_{i<k}P_{J_{ik}=J} \,,
\end{equation}
where $J_{ik}$ is the combined angular momentum of the $i$th and $k$th
quasinucleons and $P_{J_{ik}=J}$ the projection onto the $J_{ik}=J$
eigenspace. The constant $c_J$ is the interaction matrix element of
quasinucleon pairs with angular momentum $J$. It is also well known
that in the pure single-$j$-shell model these matrix elements are
equal for pairs of particles and pairs of holes. To complete
definitions, I introduce the function
\begin{equation}
  n(J,|\psi\ket)=\bra\psi|\sum_{i<k}P_{J_{ik}=J}|\psi\ket \,,
\end{equation}
which is the weight of $c_J$ in the expectation value
$\bra\psi|V|\psi\ket$, where $|\psi\ket$ is an arbitrary state in the
configuration space.

\begin{figure}
{\center\includegraphics[width=\columnwidth]{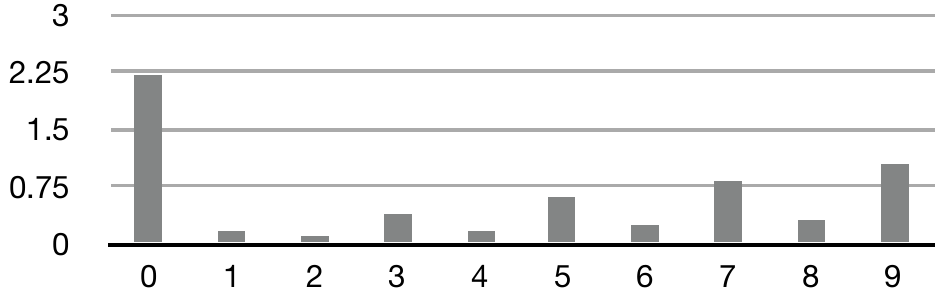}\par}
  \caption{\label{fig:nJphi} $n(J,|\phi\ket)$ as a function of $J$ for
  $j=9/2$. The total is $\sum_{i<k}\sum_J P_{J_{ik}=J}=\sum_{i<k}1=6$.}
\end{figure}

Let me summarize briefly some known facts about an interaction $V_0$.
Since it is symplectically invariant, it commutes with $s$ and $t$.
From expressions given by Edmonds and~Flowers~\cite{ref:EdFl}, it
follows that it has eigenvalue zero for $s=4$, $t=T$. Within the
$I=T=0$ space, it is therefore proportional to the projection onto the
one-dimensional $s=0$ subspace, and $|\phi\ket$ is the ground state
when a pairing interaction, that is, an attractive $V_0$, is the only
two-quasinucleon interaction. Figure~\ref{fig:nJphi}
shows the distribution of $n(J,|\phi\ket)$ for $j=9/2$. It is seen
that in spite of $|\phi\ket$ being the ground state when an attractive
$V_0$ is the only two-quasinucleon interaction, $n(0,|\phi\ket)$ makes
up only about one third of the total. The next largest
$n(J,|\phi\ket)$, making up about one sixth of the total, is
$n(2j,|\phi\ket)$. This demonstrates that an interaction which
formally favors a certain pair angular momentum does not prevent the
major part of the total weight from residing in very different angular
momenta. It can be inferred, moreover, that an additional attractive
$V_{2j}$ will help to stabilize $|\phi\ket$. It may be noticed in
passing that the broad distribution of $n(J,|\psi\ket)$ found for
$|\psi\ket=|\phi\ket$ lends some doubt to the reliability of this
distribution as an indicator of the pairing structure of any
multi-quasinucleon state $|\psi\ket$.

\section{Lowest $I=0$ state of an attractive $V_{2j}$}

The state $|\Jd=2j\ket$ is not antisymmetric in the quasineutrons or
the quasiprotons and thus violates the Pauli principle. For $j=9/2$
only 50~\% of it belongs to the even-$\Je$ space. This percentage is
typical of all $j$ and $\Jd$. The state reported by Qi\etal\ to make
up 92-95~\% of the calculated ground states is
$|\chi\ket=NP|\Jd=2j\ket$ with a normalization factor $N$ such that
$\bra\chi|\chi\ket=1$, where $P$ is the antisymmetrizer in the
quasineutrons and the quasiprotons. Quite generally,
$P|((j_1j_3)J(j_2j_4)J')I\ket$ is an exact $T=0$ state when $J$ and
$J'$ are odd, and an exact $T=1$ state when $J+J'$ is odd. Thus, in
particular, $|\chi\ket$ is an exact $T=0$ state. As shown by Moya de
Guerra\etal~\cite{ref:Mo} (also see Zamick
and~Escuderos~\cite{ref:Za}), in the $I=0$ space and for odd $J$ one
has
\begin{equation}\label{eq:VJ}
  V_J\propto P|\Jd=J\ket\bra\Jd=J|P \,,
\end{equation}
so, in particular,
\begin{equation}\label{eq:VJ}
  V_{2j}=c_J|\chi\ket\bra\chi| \,.
\end{equation}
Thus $|\chi\ket$ is the lowest $I=0$ state when an attractive $V_{2j}$
is the only two-quasinucleon interaction. (Zamick and~Escuderos
show~\cite{ref:Za} for $j=9/2$ that it is \emph{not the ground
  state}.)

\begin{figure}
{\center\includegraphics[width=\columnwidth]{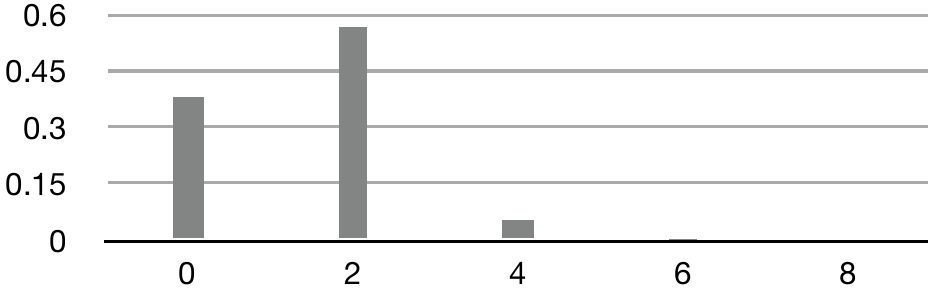}\par}
  \caption{\label{fig:2j} $\Je$ distribution in $|\chi\ket$ for
    $j=9/2$.}
\end{figure}

\begin{figure}
{\center\includegraphics[width=\columnwidth]{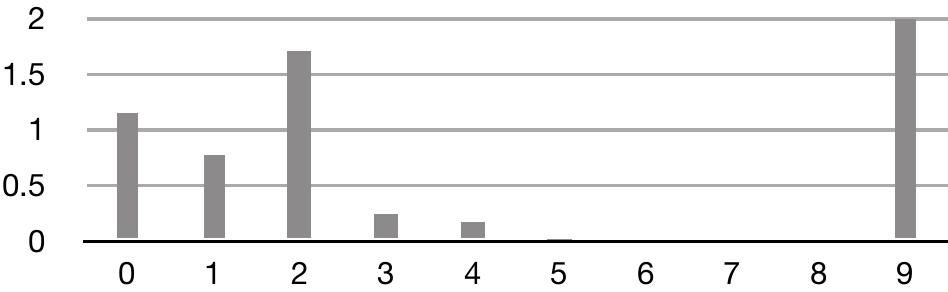}\par}
  \caption{\label{fig:nJchi} $n(J,|\chi\ket)$ as a function of $J$ for
    $j=9/2$.}
\end{figure}

The $\Je$ distribution in the state $|\chi\ket$ is shown for $j=9/2$
in Figure~\ref{fig:2j}.
Like that of $|\phi\ket$ it is dominated by low $\Je$ while the high
$\Je$ are virtually absent. A difference is that in $|\chi\ket$ the
components with $\Je=0$ and 2 contribute almost equally with some
predominance of the latter. A similar picture emerges for all $j$ with
$\Je=2$ contributing more than $\Je=0$ for $j\ge 7/2$.
Figure~\ref{fig:nJchi}
shows the distribution of $n(J,|\chi\ket)$ for $j=9/2$. Almost exactly
one third of the total is contributed by $n(2j,|\chi\ket)$, while
predominantly low $J$ give the rest. This distribution is thus quite
similar to that of $|\phi\ket$ with the low and high $J$ interchanged,
and one can infer that an attractive $V_0$ will help to stabilize
$|\chi\ket$.

\begin{figure}
{\center\includegraphics[width=\columnwidth]{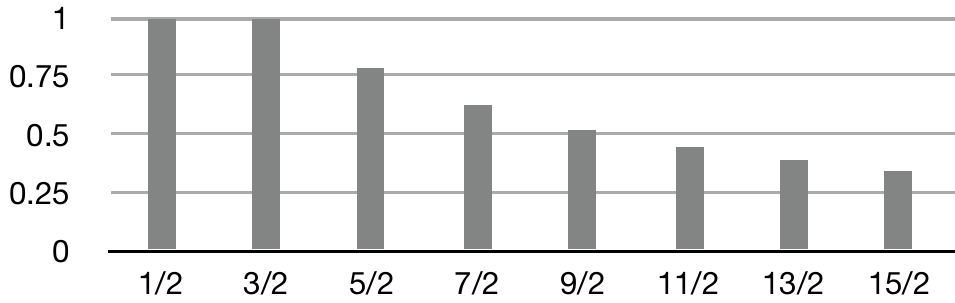}\par}
 \caption{\label{fig:olp}$|\bra\phi|\chi\ket|^2$ as a function of
    $j$.}
\end{figure}

The overlap $|\bra\phi|\chi\ket|^2$ is shown in Figure~\ref{fig:olp}
as a function of $j$. For $j\le3/2$ this overlap is one because there
is only one $I=T=0$ state. For higher $j$ it decreases with $j$, but
far from becoming orthogonal, $|\phi\ket$ and $|\chi\ket$
\emph{maintain a considerable overlap} for all relevant $j$. For
$j=9/2$ it is 52~\%.

\section{Ground state}

To show what this implies for the ground state structure I turn to
calculations for the $1f_{7/2}$ and $1g_{9/2}$ shells with effective
interactions from the literature. In both shells the $I=T=0$ space is
three-dimensional, so I introduce a third basic vector $|\xi\ket$
orthogonal to both $|\phi\ket$ and $|\chi\ket$. The states
$|\phi\ket$, $|\chi\ket$, and $|\xi\ket$ span the $I=T=0$ space, but
as already noted, $|\phi\ket$ and $|\chi\ket$ are not orthogonal. The
results of calculations are shown in Table~\ref{tab:res}
\begin{table}
  \caption{\label{tab:res}
    Composition of $|\psi\ket$ for various interactions.}
\begin{ruledtabular}
\begin{tabular}{lccc}
& $|\bra\phi|\psi\ket|^2$ & $|\bra\chi|\psi\ket|^2$
& $|\bra\xi|\psi\ket|^2$ \\
\hline
1f$_{7/2}$ \\ [3pt]
$V_0$ & 1 & 0.62 & 0 \\
$V_{2j}$ & 0.62 & 1 & 0 \\ [3pt]
SchTr, emp. & 0.78 & 0.97 & $1.5\times10^{-3}$ \\
SchTr, fit & 0.82 & 0.95 & $7\times10^{-4}$ \\
ZR I& 0.83 & 0.95 & $3\times10^{-4}$ \\
ZR II& 0.76 & 0.98 & $1.7\times10^{-5}$ \\ [5pt]
1g$_{9/2}$ \\ [3pt]
$V_0$ & 1 & 0.52 & 0 \\
$V_{2j}$ & 0.52 & 1 & 0 \\ [3pt]
SchTr, emp. & 0.77 & 0.93 & $1.4\times10^{-3}$ \\
SchTr, fit & 0.77 & 0.93 & $2.4\times10^{-3}$ \\ 
QLW & 0.78 & 0.92 & $1.1\times10^{-4}$ \\
ZE I & 0.88 & 0.84 & $1.7\times10^{-3}$ \\
ZE II & 0.79 & 0.92 & $9\times10^{-4}$ \\
ZE III & 0.89 & 0.83 & $9\times10^{-4}$ \\
ZE IV & 0.83 & 0.89 & $3\times10^{-6}$ \\
CCGI & 0.82 & 0.89 & $4\times10^{-3}$ \\
SLGT0 & 0.80 & 0.91 & $1.4\times10^{-3}$ \\
GF & 0.76 & 0.93 & $1.3\times10^{-3}$ \\
Nb90ZI & 0.77 & 0.93 & $1.5\times10^{-3}$ \\
\end{tabular}
\end{ruledtabular}
\end{table}
with the ground state denoted by $|\psi\ket$. Those for an attractive
interaction $V_0$ or $V_{2j}$ can be inferred from the preceding
discussion but are repeated for reference. Otherwise the interactions
SchTr are taken from the appendix of the aforesaid study by Schiffer
and~True with `emp.'\ referring to the empirical matrix elements and
`fit' to those derived from a universal interaction fitted to the
data. The interactions ZR~I and~II are Models~I and~II of Zamick and
Robinson~\cite{ref:ZR}. They were derived from the spectra of
$^{42}$Sc and~$^{54}$Co, respectively. QLW is $0g_{9/2}$ of Qi,
Liotta, and~Wyss~\cite{ref:QLW}. It was extracted from an interaction
for the $2p_{1/2}+1g_{9/2}$ configuration space provided by Johnstone
and Skouras~\cite{ref:JS}. ZE~I--IV are from Zamick
and~Escuderos~\cite{ref:ZE}. Specifically, ZE~I and~II are their
interactions INTc and INTd. The former consists of a $T=1$ part from
the spectrum of $^{98}$Cd and a $T=0$ part from a delta interaction.
The latter has a lower $c_{2j}$. ZE~III and~IV are from the spectrum
of $^{90}$Nb with different choices of $1^+$ level. CCGI is adapted
from the $V_{\text{low-}k}$ of Coraggio, Covello, Gargano, and
Itaco~\cite{ref:CCGI}. This is not isobarically invariant. To preserve
isospin conservation, I have used their neutron-proton matrix elements
in all channels. SLGT0, GF, and Nb90ZI are from Zerguine and~Van
Isacker~\cite{ref:ZI}. Specifically, SLGT0 and GF were constructed by
renormalization to the $1g_{9/2}$ subspace of interactions for the
$2p_{1/2}+1g_{9/2}$ configuration space provided, respectively, by
Serduke, Lawson, and Gloeckner~\cite{ref:SLGT0}, and Gross and
Frenkel~\cite{ref:GF}. Finally, Nb90ZI is from the spectrum of
$^{90}$Nb with yet another choice of $1^+$ level.

By construction, an attractive $V_0$ gives $|\bra\phi|\psi\ket|^2=1$
and $|\bra\chi|\psi\ket|^2=|\bra\phi|\chi\ket|^2$, and oppositely for
an attractive $V_{2j}$. Referring to the conclusion of Qi\etal\ cited
in the introduction, $|\bra\chi|\psi\ket|^2$ is, indeed, mostly larger
than $|\bra\phi|\psi\ket|^2$. However, \emph{both} are close to one
and considerably larger than $|\bra\phi|\chi\ket|^2$. The expectation
of a \emph{cooperation} of the terms $V_0$ and $V_{2j}$ in the
effective interaction (more precisely of their matrix elements
relative to most of the rest), is thus borne out remarkably. The
overlaps shown in the table result from expansions
$|\psi\ket\propto|\phi\ket+\alpha|\chi\ket+\beta|\xi\ket$ with
positive $\bra\phi|\chi\ket$ and $\alpha=0.9$--3.3. The ratio $\beta$
is small due to a subtle cancellation of the individual matrix
elements $\bra\phi|V_J|\xi\ket$ or $\bra\chi|V_J|\xi\ket$ for
$1\le J\le2j-1$ comparable to $c_J$.

Most notably $|\bra\chi|\psi\ket|^2$ is consistently \emph{less} in
the $1g_{9/2}$ than in the $1f_{7/2}$ shell. Thus large values of
$|\bra\chi|\psi\ket|^2$ do not arise specifically from the nucleus
approaching $^{100}$Sn. On the contrary, $|\psi\ket$ has more the
character of $|\chi\ket$ in the maximal-$j$ subshells of the
\emph{lower} major shells than in those of the higher ones. (As
already noted, in the $1p_{3/2}$ shell the state $|\chi\ket$ would
make up 100~\% of a hypothetical $s=T=0$ ground state of $^{8}$Be. In
the $1s_{1/2}$ shell a two-quasineutron-two-quasiproton state is just
a full or empty shell. Both these states can be written in the form
$|\phi\ket$ as well as in the form $|\chi\ket$.) The overlap
$|\bra\phi|\psi\ket|^2$ is seen to be about equal in both shells,
roughly 80~\%. Thus, to the extend that $|\phi\ket$, which is the sole
state in the configuration space with $s=T=0$, is considered a
manifestation of conventional isovector pairing, the latter is far
from absent from the calculated ground states.

To examine the role of the terms in the effective interaction other
than $V_0$ and $V_{2j}$, I consider an interaction $V$ with $c_J=c$
for all $1\le J\le 2j-1$, where $c$ is constant. Then
\begin{equation}
  |\psi\ket\propto|\phi\ket+\alpha|\chi\ket\
\end{equation}
with
\begin{gather}
  \alpha=\sqrt{r+a^2}+a \,, \\
  r=\frac{(c-c_{2j})n(2j,|\chi\ket)}{(c-c_0)n(0,|\phi\ket)} \,, \\
  a=\frac{(r-1)}{2|\bra\phi|\chi\ket|} \,.
\end{gather}
Keeping $c_0$ and $c_{2j}$ and setting $c=2.5$ MeV for the
interactions ZR~I--II, QLW, and ZE~I--II, which have $c_0=0$, and
$c=0$ for the rest gives $\alpha=0.3$--1.4. This is seen to be quite
different from what is calculated with the actual sets of matrix
elements. The variation of $c_J$ for $1\le J\le2j-1$ thus
significantly influences the ratio $\alpha$. As can be inferred from
Figure~\ref{fig:nJchi}, mainly $c_1$ and $c_2$ being generally less
than the average in this interval is responsible for shifting $\alpha$
towards larger values.

\section{Summary }

I studied the ground states of nuclei with two neutrons and two
protons or two neutron holes and two proton holes in a single
$j$-shell. I found that the $s=T=0$ state $|\phi\ket$, where $s$ is
the seniority and $T$ the isospin, and the lowest angular momentum
$I=0$ state $|\chi\ket$ produced by an attractive interaction of
quasinucleon pairs with angular momentum $2j$ have a large overlap for
all relevant $j$. Moreover, $n(0,|\phi\ket)$ and $n(2j,|\chi\ket)$
contribute only about one third of the total distribution of
$n(J,|\phi\ket)$ or $n(J,|\chi\ket)$, where, for any
multi-quasinucleon state $|\psi\ket$, the symbol $n(J,|\psi\ket)$
denotes the weight of the matrix element $c_J$ of the effective
interaction of quasinucleon pairs with angular momentum $J$ in the
expectation value of the total interaction. The weights
$n(2j,|\phi\ket)$ and $n(0,|\chi\ket)$ contribute about one sixth of
the total distributions. In the $1f_{7/2}$ and $1g_{9/2}$ shells, the
large negative values of $c_0$ and $c_{2j}$ relative to most of the
other $c_J$ therefore \emph{cooperate} to produce a ground state
$|\psi\ket$ which is essentially a linear combination of $|\phi\ket$
and $|\chi\ket$ with comparable coefficients. The interaction matrix
elements in channels other than $J=0$ and $J=2j$, mainly $J=1$ and
$J=2$, contribute significantly to enhance the component of
$|\psi\ket$ proportional to $|\chi\ket$. The overlap
$|\bra\phi|\psi\ket|^2$ is about 80\% in both shells. The overlap
$|\bra\chi|\psi\ket|^2$ has values 95--98~\% in the $1f_{7/2}$ shell
and 83--93~\% in the $1g_{9/2}$ shell for the interactions from the
literature that were studied. It thus \emph{decreases} as the double
magic nucleus $^{100}$Sn is approached.

\bibliography{jshell}

\end{document}